\documentclass{aa}  
\usepackage{graphicx}
\usepackage{txfonts}

\usepackage{lscape}
\usepackage{amssymb}
\usepackage{natbib}  
\bibpunct{(}{)}{;}{a}{,}{,} 
\usepackage{longtable}
\usepackage{graphicx}
\usepackage[hang,bf,footnotesize]{subfigure}
\usepackage[figuresright]{rotating}

\begin{document}
\titlerunning{Disk-bearing stars observed in scattered light}	
   \title{Pre-main sequence stars with disks in the Eagle Nebula observed in scattered light.}


   \author{M. G. Guarcello\inst{1,2,3} \and F. Damiani\inst{2} \and G. Micela\inst{2} \and G. Peres\inst{1} \and L. Prisinzano\inst{2}     \and S. Sciortino\inst{2}}

   \offprints{mguarce@astropa.unipa.it}

   \institute{Dipartimento di Scienze Fisiche ed Astronomiche, Universit\'a di Palermo, Piazza del Parlamento 1, I-90134 Palermo Italy
   \and
   INAF - Osservatorio Astronomico di Palermo, Piazza del Parlamento 1, 90134 Palermo Italy
   \and
   Actually at Smithsonian Astrophysical Observatory, MS-3, 60 Garden Street, Cambridge, MA 02138, USA}

  \date{}
 
  \abstract
   {NGC~6611 and its parental cloud, the Eagle Nebula (M16), are well-studied star-forming regions, thanks to their large content of both OB stars and stars with disks and the observed ongoing star formation. In our previous studies of the Eagle Nebula, we identified 834 disk-bearing stars associated with the cloud, after detecting their excesses in NIR bands from $J$ band to 8.0$\mu m$. }
   {In this paper, we study in detail the nature of a subsample of disk-bearing stars that show peculiar characteristics. They appear older than the other members in the $V$ vs. $V-I$ diagram, and/or they have one or more IRAC colors at pure photospheric values, despite showing NIR excesses, when optical and infrared colors are compared.}
   {We confirm the membership of these stars to M16 by a spectroscopic analysis. The physical properties of these stars with disks are studied by comparing their spectral energy distributions (SEDs) with the SEDs predicted by models of T-Tauri stars with disks and envelopes.}
   {We show that the age of these stars estimated from the $V$ vs. $V-I$ diagram is unreliable since their $V-I$ colors are altered by the light scattered by the disk into the line of sight. Only in a few cases their SEDs are compatible with models with excesses in $V$ band caused by optical veiling. Candidate members with disks and photospheric IRAC colors are selected by the used NIR disk diagnostic, which is sensitive to moderate excesses, such as those produced by disks with low masses. In 1/3 of these cases, scattering of stellar flux by the disks can also be invoked.}
   {The photospheric light scattered by the disk grains into the line of sight can affect the derivation of physical parameters of Class~II stars from photometric optical and NIR data. Besides, the disks diagnostic we defined are useful for selecting stars with disks, even those with moderate excesses or whose optical colors are altered by veiling or photospheric scattered light.} 


   \maketitle
%

\section{Introduction}
\label{intro}
	The evolution and the physical properties of circumstellar disks around young pre-main sequence (PMS) stars is currently a hot topic of astrophysics. These objects, in fact, provide the environment where planetary formation takes place, and we are far from fully understanding the mechanisms and conditions that make the disks evolve toward planetary systems. \par
With current telescopes, circumstellar disks can be spatially resolved only in very nearby star-forming regions (less distant than a few hundred parsecs). In more distant targets, photometric observations allow selecting a large number of PMS stars with disks by detecting their infrared excesses, with respect to the photospheric emission, from the disk itself. In these cases, the physical parameters of disks can be studied through the analysis of their infrared emission using models of stars with disks, and possibly outer envelope. In particular, the observed spectral energy distributions (SEDs), i.e. the fluxes emitted by the sources in various photometric bands vs. their central wavelengths, can be compared to models predicting SEDs, in order to estimate the physical properties of the systems. This is an indirect method of studying the parameters of stars with disks, strongly depending on the reliability of the chosen models. However, in some cases the estimates of the physical parameters of accreting disk-bearing stars obtained with SED analysis has been successfully compared with the values derived with other independent methods (e.g. \citealt{Robi07}).\par
	Since the publication of the initial models of astrophysical accretion disks (i.e. \citealp{SS78}, and \citealp{Prin81}), several authors developed  models of circumstellar disks around PMS stars. Recently, a detailed model of the structure of these disks has been developed by \citet{dale98,dale99,dale01}. A full summary of the physical characteristics and the simplifying assumptions of this model can be found in \citet{dale05}. The available model grid comprises 18 different central stars with four disk models each (characterized by four different accretion rates). \par
More recently, \citet{Robi06} have presented a set of models that covers an extensive parameters space: 20\,000 young stellar objects (YSO) models, each at 10 different inclination angles, for a total of 200\,000 distinct SEDs. These models take the emission of the collapsing envelope into account and allow separating the photospheric light directly observed from the light scattered by the disk into the line of sight. This feature of these models is crucial for this work. \par
	In this paper we analyze the SEDs of a subsample of candidate disk-bearing stars associated to the Eagle Nebula (M16). This star-forming region, and the associated cluster NGC~6611, has been the subject of our precedent studies:  \citet{io07}, hereafter GPM07, \citet{Io09}, hereafter GDM09, and \citet{Io10}, hereafter GMP10. This paper is organized as follows. In Sect. \ref{clupar} we summarize the parameters of NGC~6611 and M16 found in our preceding studies; in Sect. \ref{ddiag} we discuss the disk diagnostics used in our previous papers; in Sect. \ref{bludisk} we analyze and discuss the SEDs of candidate members with disks that in the $V$ vs. $V-I$ diagram appear older than the other cluster members; in Sect. \ref{evolved} we study the SEDs of the candidate members with disks with one or more IRAC colors at photospheric values; the final results are summarized in \ref{thatsallfolks}.\par

\section{Cluster parameters and analyzed data}
\label{clupar}

	In our previous studies of the young open cluster NGC~6611 and its parental cloud M16, we compiled a multiband catalog of this region including 190684 sources, using
\begin{itemize}
\item optical data in BVI bands obtained with the Wield Field Camera (WFI) at ESO/2.2m from observations which are part of the ESO Imaging Survey (\citealp{Moma01}) and which have been reduced and analyzed in GPM07;
\item NIR data in JHK bands obtained from the 2 Micron All Sky Survey (2MASS, \citealp{Cutri03}) Point Source Catalog, analyzed in GPM07;
\item NIR data in JHK bands taken from the Galactic Plane Survey (\citealp{Luca08}) of the United Kingdom Infrared Deep Sky Survey (UKIDSS) analyzed in GMP10;
\item Spitzer/IRAC data in four bands centered on 3.6$\mu$m, 4.5$\mu$m, 5.8$\mu$m and 8.0$\mu$m obtained from the Galactic Legacy Infrared Mid-Plane Survey Extraordinaire (GLIMPSE; \citealt{Ben03}), analyzed in GDM09;
\item X-ray data obtained from three Chandra/ACIS-I observations: one with 78$\,$Ksec of exposure time (P.I. Linsky) centered on NGC~6611 and two of 80$\,$Ksec, analyzed in GMP10 (P.I. Guarcello), centered on the pillar of gas called ``Column$\,$V'' \citep{Oli08} and on a  dense region of the molecular cloud.
\end{itemize}

	The WFI, 2MASS, and UKIDSS data are available in a field of $33^{\prime}\times34^{\prime}$ centered on NGC~6611 ($15.8 \times 16.2$ parsec at the distance of the cloud); also the IRAC data are available in this field, with the exception of a region northwest of about  $7^{\prime}\times10^{\prime}$. X-ray data cover about 3/4 of this studied field. Data at even longer wavelengths of M16 are provided by the MIPSGAL survey \citep{Car09}, but the observation is affected by an intense and patchy background \citep{Fla09} in correspondence with NGC6611, which complicates the estimation of point sources flux.  For these reasons, these data are not studied in the present work. \par
In our previous works we evaluated several physical properties of M16 and its stellar population, such as distance, age spread, and average extinction. In these works, the influence of the OB members of NGC~6611 on the evolution of circumstellar disks and the star formation in the nebula was also studied. Table \ref{param} summarizes some of the physical properties of NGC~6611 and M16 that we found. See \citet{Oli08} for a complete review of the Eagle Nebula. \par

	\begin{table}[]
	\centering
	\caption {Physical properties of NGC~6611 and M16 obtained in GPM07, GDM08, and GMP10.}
	\vspace{0.5cm}
	\begin{tabular}{cc}
	\hline
	\hline
	Parameter& Value \\
	\hline
	Distance					&1750 parsec\\
	Median age of NGC~6611				&$\sim 1$ Myear\\
	Age range of stars in M16			&$\sim3 - 0.3$ Myears \\
	Average DF$^*$ in NGC~6611			&$36\% \pm 1 \%$ \\
	Average DF$^*$	far away from OB stars		&$42\% \pm 3 \%$ \\
	Average DF$^*$	close to OB stars		&$28\% \pm 3 \%$ \\
	Average $A_V$	in NGC~6611			&2.6$^m$   \\
	$R_V$						&3.3   \\
	Core relaxation time				&5.2 Myears\\
	\hline
	\hline
	\multicolumn{2}{l}{$^*$DF stands for Disk Frequency} \\
	\end{tabular}
	\label{param}
	\end{table}

In our previous studies we identified 834 candidate disk-bearing stars. Among them, there are several stars with interesting color peculiarities (discussed in the following sections). In this paper we study their properties using the SED fitting tool\footnote{available at http://caravan.astro.wisc.edu/protostars/index.php} developed by \citet{Robi06}. In our analysis, for each studied star, we select the models that best reproduce the observed SEDs, using the compatibility criterion defined in \citealt{Robi07} (i.e. those for which the reduced chi-square satisfies the relation $\chi^2-\chi_{best}^2\leq3$, where $\chi_{best}^2$ is that of the best-fit model). \par

\section{Reddening-free color indices used for disk selection.}
\label{ddiag}

	To select stars with infrared excesses, in our previous studies we used a set of color indices ($Q$ indices), which are defined to be independent of reddening \citep{Dami06}. The set is composed of nine $Q$ reddening-free color indices, defined with the bands used in our previous studies. Eight indices compare the infrared colors with $V-I$, assumed to be representative of photospheric emission:

	\begin{equation}
	Q_{VIAB} = \left( V-I \right) - \left( A-B \right) \times E_{V-I}/E_{A-B}
	\label{Q1def}
	\end{equation}

	and one compares $J-H$ with $H-K$, i.e.. \par

	\begin{equation}
	Q_{JHHK} = \left( J-H \right) - \left( H-K \right) \times E_{J-H}/E_{H-K}
	\label{Q2def}
	\end{equation}
	
In these equations, $A-B$ is an infrared color ($I-J$, $J-H$, $J-K$, $H-K$, $J-[3.6]$, $J-[4.5]$, $J-[5.8]$ and $J-[8.0]$); $E_{V-I}$, $E_{J-H}$, $E_{H-K}$, and $E_{A-B}$ are reddening in the corresponding colors, which have been calculated from the reddening laws of \citet{RL85}, \citet{Mat90}, and \citet{Muna96}. Hereafter, we refer to the whole list of indices with $Q$ or $Q_{ABCD}$, and call $Q_{2MASS}$ and $Q_{UKIDSS}$ the indices defined with $VIJHK$ photometry and IR data respectively from 2MASS/PSC and UKIDSS/GPS, while $Q_{IRAC}$ and $Q_{VIJ[sp]}$ are those involving IRAC bands. \par

	\begin{figure}[]
	\centering	
	\includegraphics[width=7cm]{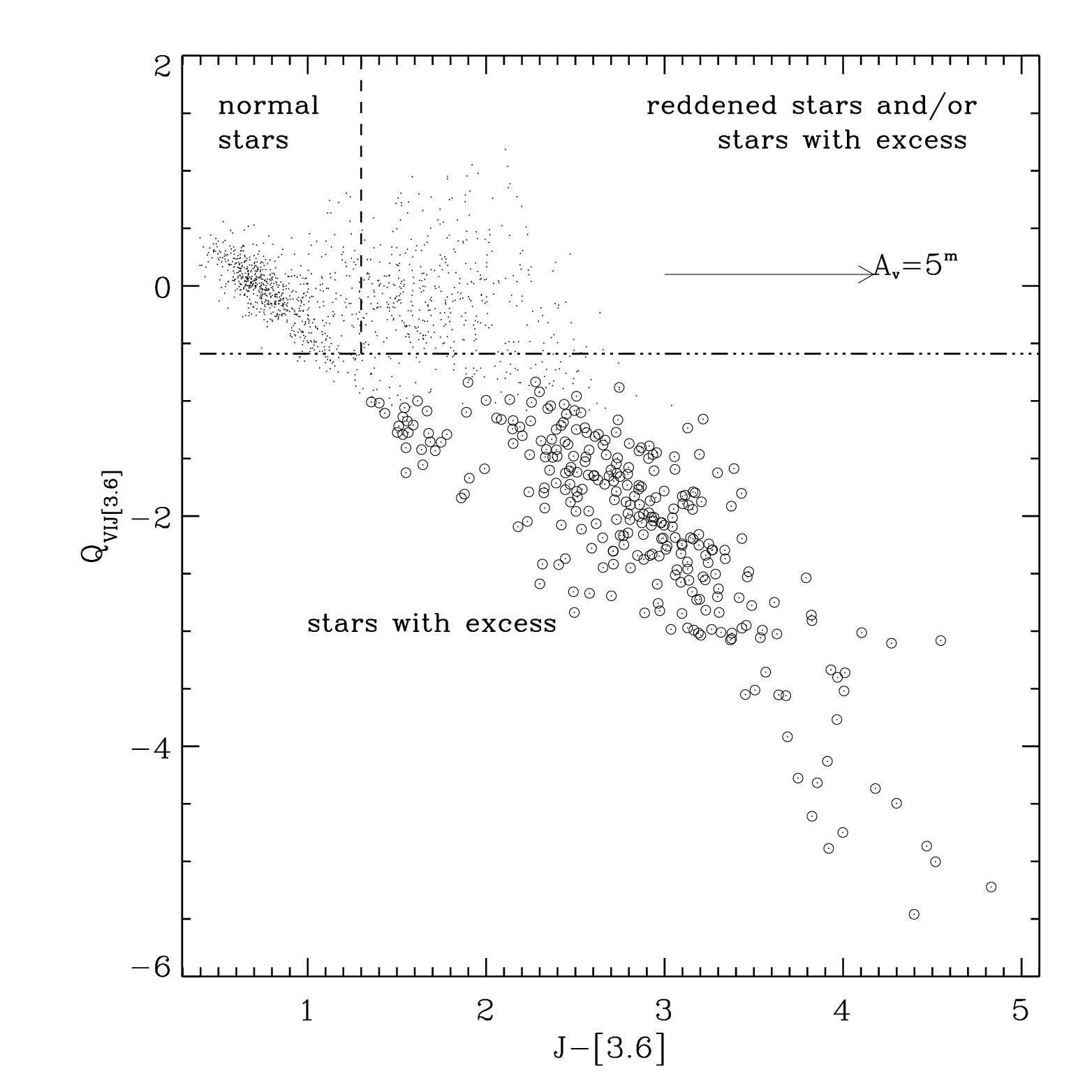}
	\caption{$Q_{VIJ[3.6]}$ vs. $J-[3.6]$ diagram of the stars in the studied field (points). The loci of normal stars, stars with excesses in [3.6] and that of stars for which the effects of reddening and excesses cannot be discerned are shown. Stars with excesses in [3.6] are marked by circles. The reddening vector is also shown.}
	\label{Qesfig}
	\end{figure}

Figure \ref{Qesfig} illustrates the case of the $Q_{VIJ[3.6]}$ vs. $J-[3.6]$ diagram used to select stars with excesses in [3.6] in GDM09. In this figure are plotted both the color and index of all the stars falling in the surveyed field with $\sigma_{V-I}$ and $\sigma_{J-[3.6]}$ smaller than $0.15^m$. The reddening vector is horizontal since the $Q$ index is independent of extinction. For this reason, in the $Q_{ABCD}$ diagrams, like that shown in Fig. \ref{Qesfig}, reddening shifts the star's positions rightward along the $x$ direction, from the locus of ``normal stars'' to the locus of non-classifiable stars ({\it reddened stars and/or stars with excesses}). In Fig. \ref{Qesfig}, if $J-[3.6]$ is redder than the photospheric values since there is an excess in [3.6], the index decreases while $J-[3.6]$ increases, shifting stars downward to the locus of stars with excesses. In this way, stars with excesses and reddened photospheres can often be separated, even if a number of stars remain unclassified since they fall in the locus where we cannot distinguish between reddening or excesses. The use of 9 $Q$ indices to detect excesses in every single infrared band reduces the number of unclassifiable stars. In the context of this paper, it is important to note that the indices decrease even if $V-I$ decreases. This can also select stars having $V-I$ bluer than the photospheric values (such as when there is intense optical veiling). \par
	For each $Q_{ABCD}$ index, the stars with excess in $D$ band are defined as those with the index significantly (i.e. by more than 3 times the error in the $Q$ index) smaller than the lower limit of photospheric $Q$ values (the horizontal dotted-dashed line in Fig. \ref{Qesfig}). These limits have been defined as the lower boundaries of the loci, in these diagrams, of the stars with normal colors. For $Q_{2MASS}$ and $Q_{UKIDSS}$, the locus of normal stars has been defined by using the photospheric colors obtained from the isochrones of \citet{Sie00}, with age ranging from 0.25 Myears to $10^9$ yrs and the masses ranging from 0.2-7$\,M_{\odot}$ in order to consider all possible observed disk-less stars. The isochrones of \citet{Sie00} do not provide IRAC colors, so for $Q_{IRAC}$ the locus of normal stars have been defined using the sources with photospheric IRAC colors and optical emission. GDM09 have shown that the completeness of the selection of candidate stars with disks made with $Q$ indices depends on both the mass and age of the disk-bearing stars. \par

\section{Blue stars with excesses}
\label{bludisk}

	\begin{figure*}[]
	\centering	
	\includegraphics[width=13cm]{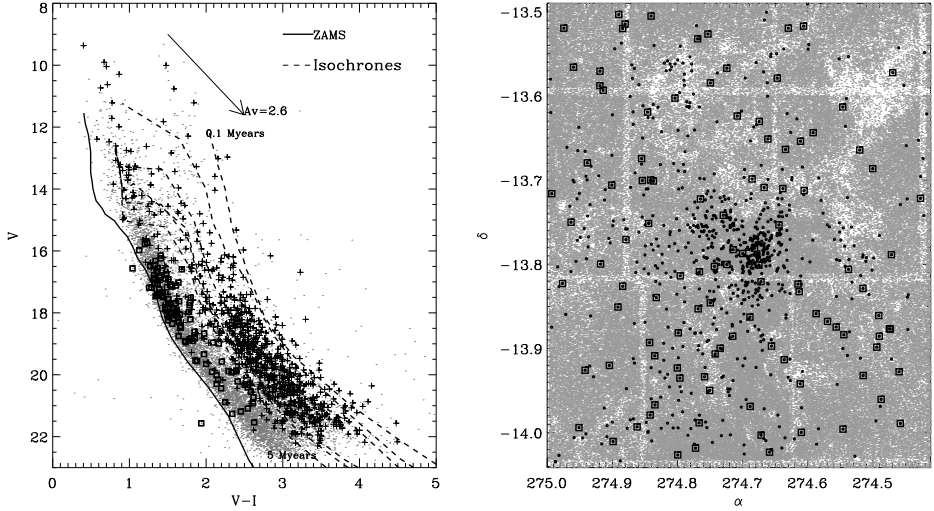}
	\caption{Left panel: $V$ vs. $V-I$ diagram of all stars in the studied FOV (gray points), X-ray sources (crosses), and BWE stars (boxes). The thick solid line is the ZAMS (from \citealt{Sie00}) at the distance of 1750~pc and with $A_V=1.45^m$. The dashed lines are the isochrones at 0.1, 0.25, 1, 2.5, 3, and 5 Myrs, with the average extinction appropriate for cluster members ($A_V=2.6^m$). The extinction vector is obtained from the law of \citet{Muna96}. Right panel: spatial distribution of stars in the multiband catalog of M16 (gray points), candidate disk-bearing stars (black dots), and BWE stars (boxes). North is up, east is on the left.}
	\label{vviblu}
	\end{figure*}
	
The left panel of Fig. \ref{vviblu} shows the $V$ vs. $V-I$ diagram for stars in the surveyed region with errors smaller than 0.1$^m$ and 0.15$^m$ in $V$ and $V-I$ respectively. Since X-rays emission of PMS stars is much more intense than for stars in main sequence (MS) phase, in this diagram the positions of stars detected in X-rays define the locus populated by the young stars associated with M16, which is delimited by the 0.1-3 Myears isochrones (from \citealp{Sie00}). In the $V$ vs. $V-I$ diagram, however, 90 stars with NIR excesses, detected by the diagnostics described in Sect. \ref{ddiag}, have the $V-I$ bluer than the PMS locus, i.e. with positions typical of field main sequence stars older than the other cluster members. Hereafter we call these stars ``blue stars with excesses'', or with the acronym $BWE$ stars. The right panel of Fig. \ref{vviblu} shows the spatial distribution of all the stars in our multiband catalog, disk-bearing stars and the stars with disks which appear older that the cluster locus in the $V$ vs. $V-I$ diagram, falling within the $33^{\prime}\times 34^{\prime}$ region centered on NGC6611.  \par
	Before discussing the characteristics of BWE stars in detail, it is important to stress that GMP07, GDM09, and GMP10 have been conservative in defining the stars with excesses, in order to avoid the contamination of the members list as much as possible. We also stress that the following characteristics allow us to be confident about the member status of most BWE stars:

	\begin{itemize}
\item In the IRAC color-color diagram, the only 4 BWE stars with emission in all IRAC bands exhibit colors compatible with ClassII YSOs.
\item In the NIR color-magnitude and color-color diagrams all BWE stars are consistent with being cluster members, with the exception of 9 stars that have NIR colors typical of foreground sources and that are be excluded from the following analysis (since they are likely field stars).
\item In 45 BWE stars with small $JHK$ errors, the excesses have only been detected with the $Q$ indices defined with optical and infrared photometry. This does not exclude these stars being mismatches between optical and NIR catalog. However, in 30 cases also the $Q_{JHHK}$ index shows a weak $K$-band excess, although it is not significant (at $3\sigma$), independently of optical photometry. In fact, the mean value of the $Q_{JHHK}$ index for the entire sample of BWE stars is equal to -0.23, much lower than the mean $Q_{JHHK}$ of all stars in the studied FOV (equal to 0.30), pointing toward the membership of the majority of this sample of stars. 
\item Two thirds of BWE stars have infrared excesses in $JHK$ both using the UKIDSS/GPS and the 2MASS/PSC photometry.
	 \end{itemize}

	Another relevant hint about the nature of BWE stars as disk-bearing YSOs arises from the preliminary analysis of optical spectroscopic observations. We obtained observations (with a total exposure time of 6 hours) of the 20 brightest BWE stars with the intermediate resolution spectrometers GIRAFFE/FLAMES@VLT (Obs. ID: 083.C-0837; P.I.: Guarcello). These observations have been performed with the {\it HR15N} (R=17000) setup, which covers the 5200-7600$\AA$ spectral range, comprising both the $H\alpha$ $6562.8\AA$ and Lithium $6707.8\AA$ lines. Among the BWE stars observed with FLAMES, 6 have an evident Li absorption line and 13 broad and asymmetric $H\alpha$ line, thus showing spectroscopic evidence of their PMS nature (4 among these stars have both features). Among the BWE stars with accretion signatures, eight have a blueshifted absorption feature (with respect to the center of the line), which is a signature of occurring inflow of gas onto the stellar surface; two stars have an asymmetric profile of the line; three stars have both the absorption feature and the asymmetric profile. Following \citet{Kuro06}, these line profiles indicate both gas accretion onto the stars and wind from the disk surface with different values of accretion rate, mass-loss rate, and disks inclination. A full account of the spectroscopic data will be  the subject of a forthcoming paper. In the present context, they show that about 3/4 of the BWE stars are indeed young PMS stars. \par
The spatial distribution of BWE stars (right panel of Fig. \ref{vviblu}) may suggest that it is uncorrelated with the cluster. However, inside NGC~6611 BWE stars are more frequent at large distances from OB members, which are concentrated in the center of the cluster. It is possible to suppose that the physical mechanisms responsible for the anomalous optical colors of BWE stars are more prominent at large distances from the center of the cluster. This hypothesis is consistent with the smaller disk fraction close to massive stars observed in our previous studies, and it can be due to a lower incident UV radiation or to the dynamical evolution of the cluster. Moreover, our studies of the outer regions of the Eagle Nebula revealed a large number of candidate young stars associated with the outer region of M16. Most of the BWE stars, then, could be part of this widespread population of M16. \par
		
	\subsection{Hypotheses on the nature of BWE stars}
	\label{bwehypo}
	
	One possible explanation of the nature of BWE stars is that they are strong accretors, characterized by an intense veiling that alters their optical colors. The possibility that the position of accreting T-Tauri stars in the H-R diagram can be affected by veiling has been discussed by several authors. For example, \citet{Har90} studied 37 K7-M1 T-Tauri stars in the Taurus-Auriga complex, and they found that strong accretion ($\dot{M} \geq 10^{-7} M_{\odot} yr^{-1}$) affects the positions in the H-R diagrams of young stars, from which it is impossible to have an accurate estimate of their age. Stars with properties similar to our BWE stars have been observed in other star-forming regions. For example, \citet{Hille97} obtained the $V$ vs. $V-I$ diagram, down to $V\sim 22^m$, of the Orion Nebula Cluster, where a few tens of K-M cluster members are even bluer than the ZAMS at the distance of the nebula. This author hypothesized that these stars can be heavily veiled or their photometry might be contaminated by the nebula emission.\par
Another possible explanation for the observed colors of the BWE stars is that they are cluster members surrounded by a dusty reflection nebula. In this case, the extinction at $I$ and infrared bands is different from that at $V$, since the optical light is reflected by the nebula and the optical colors are very different from what is expected from a reddened photosphere. This hypothesis was proposed by \citet{Dami06} for the young open cluster NGC~6530 to explain stars with excess detected with $Q_{VIIJ}$, which, in the optical color-magnitude diagram, are apparently older than the X-ray emitting members of the cluster (i.e. like our BWE stars). To support this hypothesis, \citet{Dami06} pointed out that two of the stars in the Taurus-Auriga complex, with excess detected with $Q_{VIIJ}$, are Class~I sources surrounded by an envelope, with a faint optical counterpart. This hypothesis is also supported by the direct high spatial resolution observations of the Hubble Space Telescope of stars surrounded by a reflection nebula in the Taurus-Auriga cluster \citep{Pad99}. \par 
	It is also possible that a significant fraction of stellar radiation is scattered into the line of sight by the dust grains in the disk atmosphere. If the light emitted by the star is scattered {\it away from} the line of sight, such as while passing through the isotropic interstellar medium (ISM), the source is affected by normal interstellar extinction. Stars with disks, however, are not isotropic systems. The amount of light scattered {\it into} the line of sight by the circumstellar material can be greater than the extincted amount of radiation. Since small dust grains, with sizes $\leq 1 \mu m$, scatter optical light more efficiently than infrared radiation \citep[ and references therein]{Thro01}, if the disk surface is mostly populated by small dusty grains, the optical radiation at short wavelengths is more efficiently scattered. If a large amount of such radiation is scattered into the line of sight, the observed flux emitted by the reddened photosphere at shorter wavelengths increases, giving a blue $V-I$ color. The efficiency of this mechanism depends on the distribution of the grain sizes in the disk atmosphere (which changes during disk evolution) and on other disk properties such as the disk inclination angle and its scale height.   \par
We cannot a priori reject the hypothesis that BWE stars are older than the other PMS members. There are known examples of classical T-Tauri stars much older than the disk survival times, which are usually observed or predicted by the YSOs evolutionary scenario. For example, \citet{Pal05,Pa07} studied some lithium-depleted members of the Orion Nebula Cluster, among which they found a Classical T-Tauri star with large NIR excesses despite its age (estimated from Li abundance) equal to $\sim 20$ Myrs, and \citet{Argi07} studied MP Muscae, a T-Tauri star with disk, low accretion ($\sim 10^{-11} M_{\odot}/$yr) and an aged $\sim 17$ Myrs. \par
	Finally, it is also possible that BWE stars come from to mismatches between optical and infrared catalogs. A random coincidence between a foreground MS optical star and an infrared source, in fact, can produce a source with negative $Q$ indices and ``blue" optical colors. This is indeed a reasonable hypothesis for the 9 BWE stars with $H-K$ compatible with field stars. However, the mean value of $Q_{JHHK}$ for all BWE stars (-0.23) implies that this sample is not dominated by mismatches.  \par
 
\subsection{SED analysis of BWE stars: accretion}
\label{bweaccre}

The properties of the 81 BWE stars (all the 90 BWE stars minus those with $H-K$ color compatible with field stars) have been studied by SED analysis, as described in Sect. \ref{intro}. The explored parameters space is described in \citet{Robi06}. Among the analyzed BWE stars, 43 are compatible with at least one SED model with a good $\chi^2$ value. Mostly thanks to the small photometric errors in UKIDSS bands (GMP10), these stars are typically compatible with just a few models, and the membership of 9 among them is confirmed by our spectroscopic data. For the remaining stars, their membership status is supported by the properties described in Sect. \ref{bludisk}, but we cannot discern their nature by SED analysis.  \par 
	As explained previously, high accretion rates, and the resulting intense optical veiling, could account for the blue optical colors observed in the BWE stars. However, only 5 BWE stars have the SEDs compatible with at least one model that shows, in the optical bands, non photospheric excesses due to gas accretion (hereafter, we call these stars ``$E$-$stars$''). \par
The analysis of their SEDs suggests that $\dot{M}$ is on the average higher in the E-stars sample from all the other candidate members with a disk, nevertheless the observability of the excess in optical bands requires not {\it only} high accretion rates, but also low stellar temperatures. In fact, in stars with low effective temperature, the contrast between the photospheric emission and a possible excess in $V$ band is higher than in stars with higher effective temperature. Since the SEDs of all but one E-star are also compatible with models without these optical excesses, we conclude that the optical veiling is the correct explanation for the optical colors of BWE stars only in few cases. \par

\subsection{SED analysis of BWE stars: scattering and obscuration by the disk}
\label{bwescat}

The optical colors of BWE stars might be alternatively explained with the interaction between stellar radiation and the circumstellar disk. The models of \citet{Robi06} allow us to separate the stellar radiation that is directly observed from that scattered into the line of sight by disk's dust grains. The amount of light scattered into the line of sight can significantly alter the stellar SED (which should be a reddened photosphere at the effective stellar temperature), by increasing the observed optical flux mostly at short wavelengths. Moreover, disks seen at large inclinations can partially occult the stellar photosphere, so part of stellar radiation is totally absorbed by the circumstellar material. As a consequence, a reduced photospheric flux with unchanged colors (since they are related to the stellar temperature) is observed in these cases. \par
	Among the 43 BWE stars with good SED fitting, it predicts a large fraction of optical emission composed by scattered light in 26 stars, and in 16 cases a partial occultation of the stellar photosphere by the disk without scattering effects. In the following we discuss the results obtained from the analysis of the former 26 BWE stars. \par
\begin{figure*}[]
	\centering	
	\includegraphics[width=7cm]{6611-27750_fit.eps}
	\includegraphics[width=7cm]{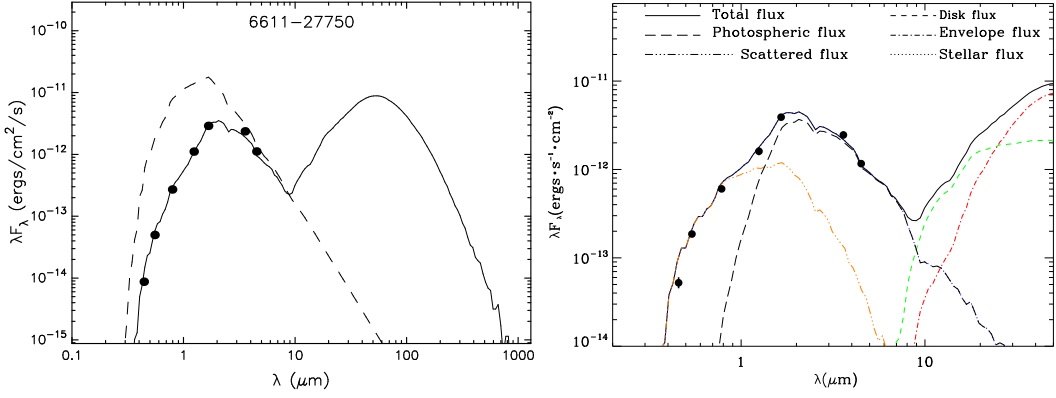}	
	\caption{Left panel: the SED model (solid lines) that reproduces the observed SED (black points) of a BWE star (6611-27750). The dashed line is the unreddened photospheric flux expected if the star should to be without disk. The right panel shows the different components of the SED model, marked with different kind of lines, as explained in the top of the panel. Black points mark the observed fluxes, which are corrected for interstellar extinction using the $A_V$ predicted by the SED model ($2.3^m$). Both models are normalized at the distance of 1 Kpc.}
	\label{sedEnoE}
	\end{figure*}

As an example, the left panel in Fig. \ref{sedEnoE} shows the SED model compatible with the observed SED of 6611-27750\footnote{This ID identifies the stars in the catalog published in GMP10}, while the right panel shows the individual components of the model. Comparing, in the right panel, the optical scattered flux with the directly observed photospheric flux, it is evident that the light scattered by the disk into the line of sight dominates the SED at optical wavelengths. Therefore, the optical SED is not simply consistent with a reddened photosphere (which is also directly observed). The scattering increases the optical flux significantly (with respect to the expected photospheric values) mostly at short wavelengths, accounting for the optical blue colors observed for this star.  \par

In the right panel of Fig. \ref {sedEnoE}, the optical part of the SED is dominated by scattered light. If we calculate, in a particular band, the fraction of observed flux that is scattered into the line of sight on to the total emitted flux as	

	\begin{equation}
	Frac_{scat}=\frac{F_{scat}}{F_{dir}+F_{scat}}
	\label{scateq}
	\end{equation}

we find that for almost all the BWE stars with significant, scattered optical light, more than $90\%$ of the observed optical flux is scattered by circumstellar material: without scattered light these stars could not have been detected in optical bands. \par
	The models of \citet{Robi06} do not provide information on the grain population in the disks (for example the average grain size), so we cannot test whether the disks in BWE stars with a large amount of scattered light are mostly populated by small grains, as expected. However, collisional aggregation and settling in the disk midplane is expected to occur during the evolution of the disks. Thanks to these phenomena, the number of small grains in the upper layers of the disk decreases during the evolution of the disk itself, thus also decreasing the amount of scattered optical light. The typical timescales for settling and aggregation are not well determined, and some models (as that developed by \citealp{DD04B}) predict short timescales of about 1 Myear for settling to occur. However, this disagrees with some older Herbig stars with disk having been observed with a significant amount of scattered light. We estimated the age of BWE stars with a significant amount of scattered flux using the SED models. The results are summarized in Table \ref{scatage}, showing that the SED analysis for almost all these stars predicts an age under 1 Myear, younger than the members median age ($\sim 1$ Myear, GMP10). This suggests that, on average, a large amount of scattered light is present among the youngest cluster members, probably associated to a large fraction of small grains in the disk upper layers, still present at young ages. However, we caution the reader not to overinterpret our finding since it is very difficult to assess whether this is a real result, or if it is produced by some assumption in the models. \citet{DD04B} list some physical effects, not considered by the available disk models, which can explain the presence of grains with typical sizes of those in the ISM in the upper layers of disks older than 1 Myear. We focus our attention just on the large difference between the age of the BWE stars predicted by the models (less than 1 Myear and compatible with the cluster age) and the age that can be deduced from the $V$ vs. $V-I$ diagram (several tens of million of years), likely caused by the scattered optical light. 
	
	\begin{table}[!h]
	\centering
	\caption {Age of the 26 BWE stars with significant amount of scattered flux predicted by SED fitting.}
	\vspace{0.3cm}
	\begin{tabular}{cc}
	\hline
	\hline
	Age(years) & $N_{stars}$ \\
	\hline
	\hline
	$\leq 10^5$      &2 \\
	$10^5\div10^6$           &21\\
	$10^6\div10^7$     	 &3 \\
	\hline
	\hline
	\multicolumn{2}{l}{} \\
	\end{tabular}
	\label{scatage}
	\end{table}

	As explained, among the 42 BWE stars for which it was possible to study their SED, in 16 cases the central star is largely occulted by the disk, and there are no evident scattering effects. In these cases, these stars are much fainter in optical band than expected from the stellar properties, while the $V-I$ is not affected by the occultation having the values that should to be expected from the stellar effective temperature. SED analysis predicts that these stars are on average older that those with a large amount of optical light scattered into the line of sight, as shown in Table \ref{scatage2}, where we report the age of these 16 stars predicted by the models.

	\begin{table}[!h]
	\centering
	\caption {Age of the 16 BWE with an obscured photosphere predicted by SED fitting.}
	\vspace{0.3cm}
	\begin{tabular}{cc}
	\hline
	\hline
	Age(years) & $N_{stars}$ \\
	\hline
	\hline
	$\leq 10^5$               &1\\
	$10^5\div10^6$           &2\\
	$10^6\div10^7$     	  &13 \\
	\hline
	\hline
	\multicolumn{2}{l}{} \\
	\end{tabular}
	\label{scatage2}
	\end{table}

\begin{figure}[!h]
	\centering	
	\includegraphics[width=5.0cm]{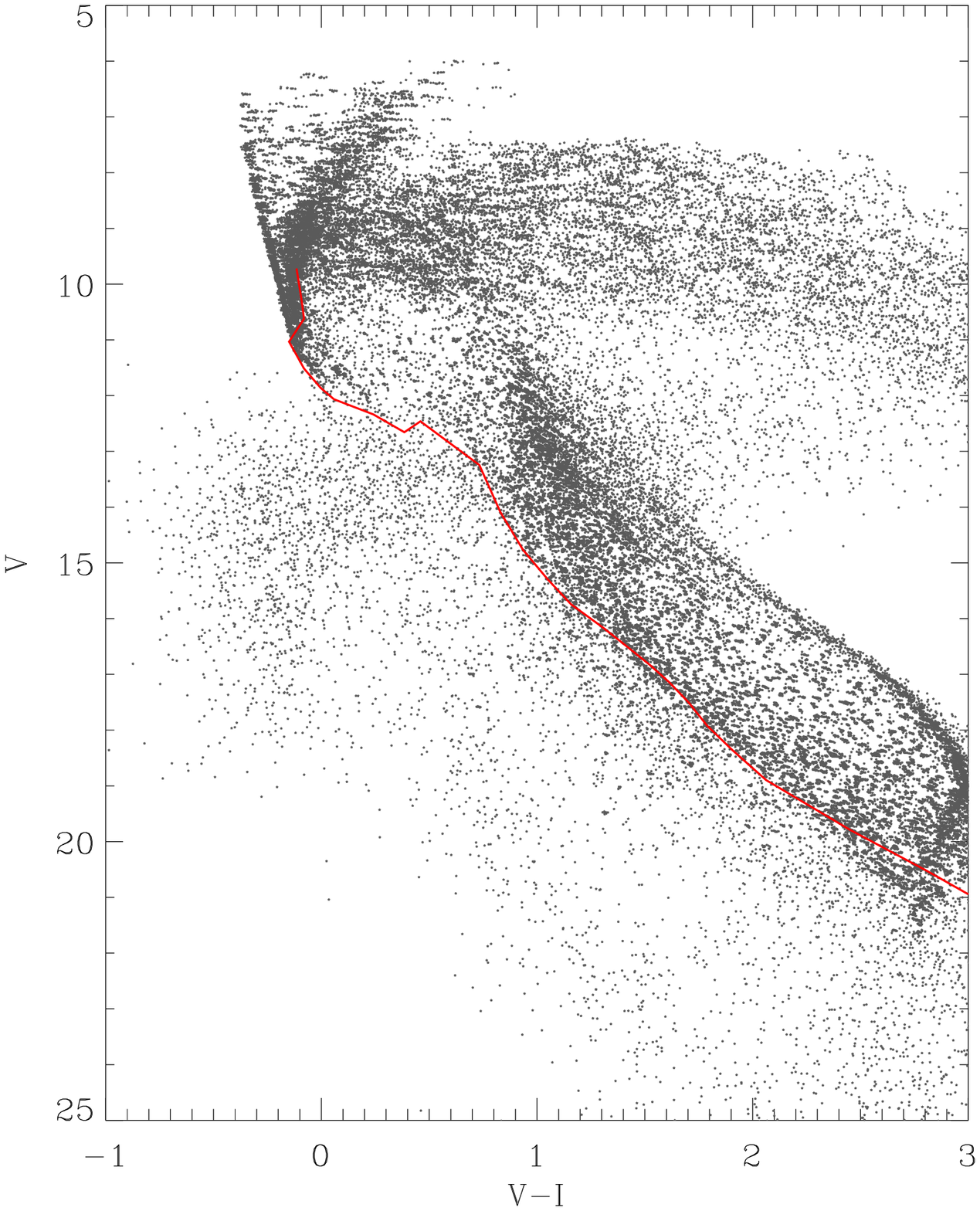}
 	\includegraphics[width=5.0cm]{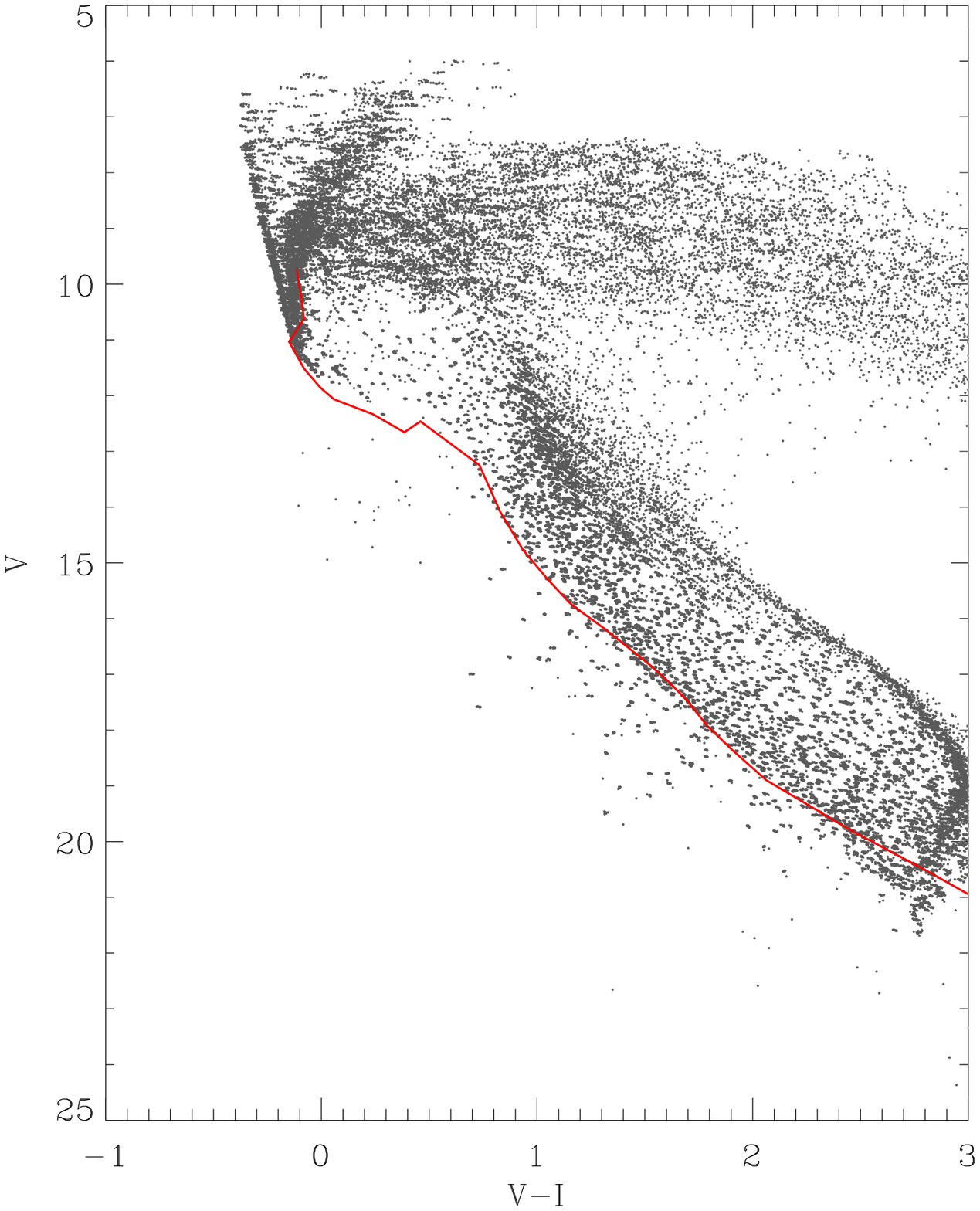}
	\caption{$V$ vs. $V-I$ diagrams obtained from the magnitudes of ClassII YSOs predicted by the models of \citet{Robi06}. The line is the 10 Myrs isochrone of \citet{Sie00}. In the upper panel all ClassII models are plotted in the lower only those with disk inclination smaller than $80^{\circ}$.}
	\label{vvirobfig}
	\end{figure}

	The SEDs of all the analyzed 42 BWE stars (both with an obscured photosphere and scattering effects), but four, are modeled with highly inclined disks (with an inclination greater than $80^{\circ}$ with respect to the line of sight). This indicates, as expected, that scattering effects are more important for disks observed at high inclination. To verify this, we produced the optical color-magnitude diagrams using all the ClassII models in the database of \citet{Robi06}. Figure \ref{vvirobfig} shows the $V$ vs. $V-I$ diagram obtained with the magnitudes of ClassII objects predicted by the models. All the available models of ClassII objects are plotted in the upper panel. They populate mostly the PMS locus, to the right of the 10 Myrs isochrone, but a number of them lie also in the locus typical of older stars, which is leftward of the 10 Myrs isochrone. Only the models for ClassII objects seen at low disk inclination angle ($\theta \le 80^{\circ}$) are plotted in the lower panel. In this diagram, points lie only to the right of the 10 Myears isochrone. The comparison between the two panels suggests that the effects related to a disk observed at high inclination (both scattering and obscuration) shift the position of the stars in the $V$ vs. $V-I$ diagram leftward of the locus of PMS stars.

\section{Candidate stars with NIR excesses and photospheric IRAC colors}
\label{evolved}

	\begin{figure}[!h]
	\centering	
	\includegraphics[width=8cm]{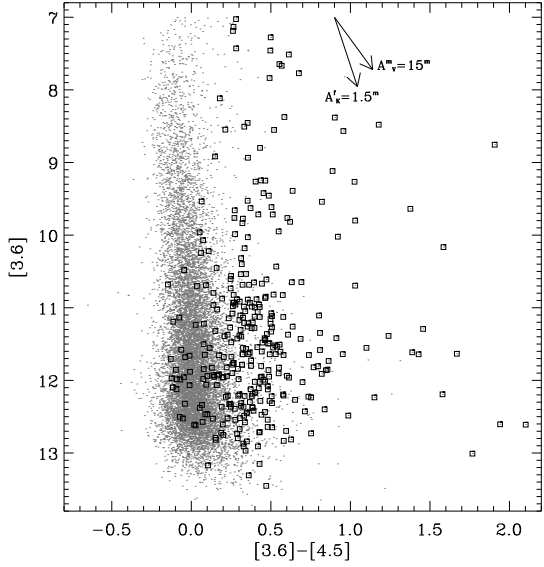}
	\caption{[3.6] vs [3.6]-[4.5] diagram of the stars in the surveyed region (small points). Boxes mark stars with NIR excesses selected with $Q$ indices. The reddening vector with A$_V$=15$^m$ was obtained from the reddening law of \citet{Mege04}, while that with A$_K$=1.5$^m$ from that of \citet{Fla07}}
	\label{colmagsp}
	\end{figure}

	Figure \ref{colmagsp} shows the [3.6] vs. [3.6]-[4.5] diagram for all the stars in our catalog with $\sigma_{[3.6]} \le 0.1^m$ and $\sigma_{[3.6]-[4.5]} \le 0.15^m$. Most candidate stars with disks lie in the region of this diagram with $[3.6]-[4.5] > 0.2^m$, showing a red color, as expected. In Figure \ref{colmagsp}, 66 stars with excesses have $[3.6]-[4.5]\leq 0.2^m$, thus showing a $[3.6]-[4.5]$ color at pure photospheric value even if infrared excesses are detected by the use of the $Q_{IRAC}$ indices. In the following, we describe why a star with disk could have such [3.6]-[4.5] colors, while having excesses in $Q_{IRAC}$ indices, and then we discern the nature of these stars studying their SEDs. \par
It is possible that the disk has an evacuated inner region, resulting in a weak emission at the NIR bands. Also a high disk inclination can self-mask the NIR emission from the inner part of the disk. However, such disks would be selected with $Q_{VIJ[8.0]}$, which can be affected by a prominent silicate band in emission. Another possibility is that the disk has a low mass, producing moderate NIR excesses with respect to the photospheric emission. In this case the [3.6]-[4.5] color is not strongly affected by the emission of disks, but $Q_{VIJ[3.6]}$ and $Q_{VIJ[4.5]}$ indices can also detect these moderate excesses since they compare $J-[3.6]$ or $J-[4.5]$ to $V-I$. This depends on the spectral type of the central star, as shown by GDM09. \par
	Among these 66 stars, only in 33 cases the physical disk parameters are determined well with SED analysis. The models compatible with the SEDs of 22 among these stars range over all the hypotheses listed above. A fraction of these stars (10) have SEDs compatible with disks having masses $\leq 10^{-5} M_{\odot}$, accounting for moderate excesses with respect to the photospheric emission at [3.6] and [4.5]. It is remarkable that $Q$ indices, since directly comparing NIR colors with $V-I$, can be effectively used to select these disks. The SEDs of the remaining 12 stars predict an almost evacuated inner disk with a prominent silicate emission band. \par
\begin{figure}[]
	\centering	
	\includegraphics[width=5cm]{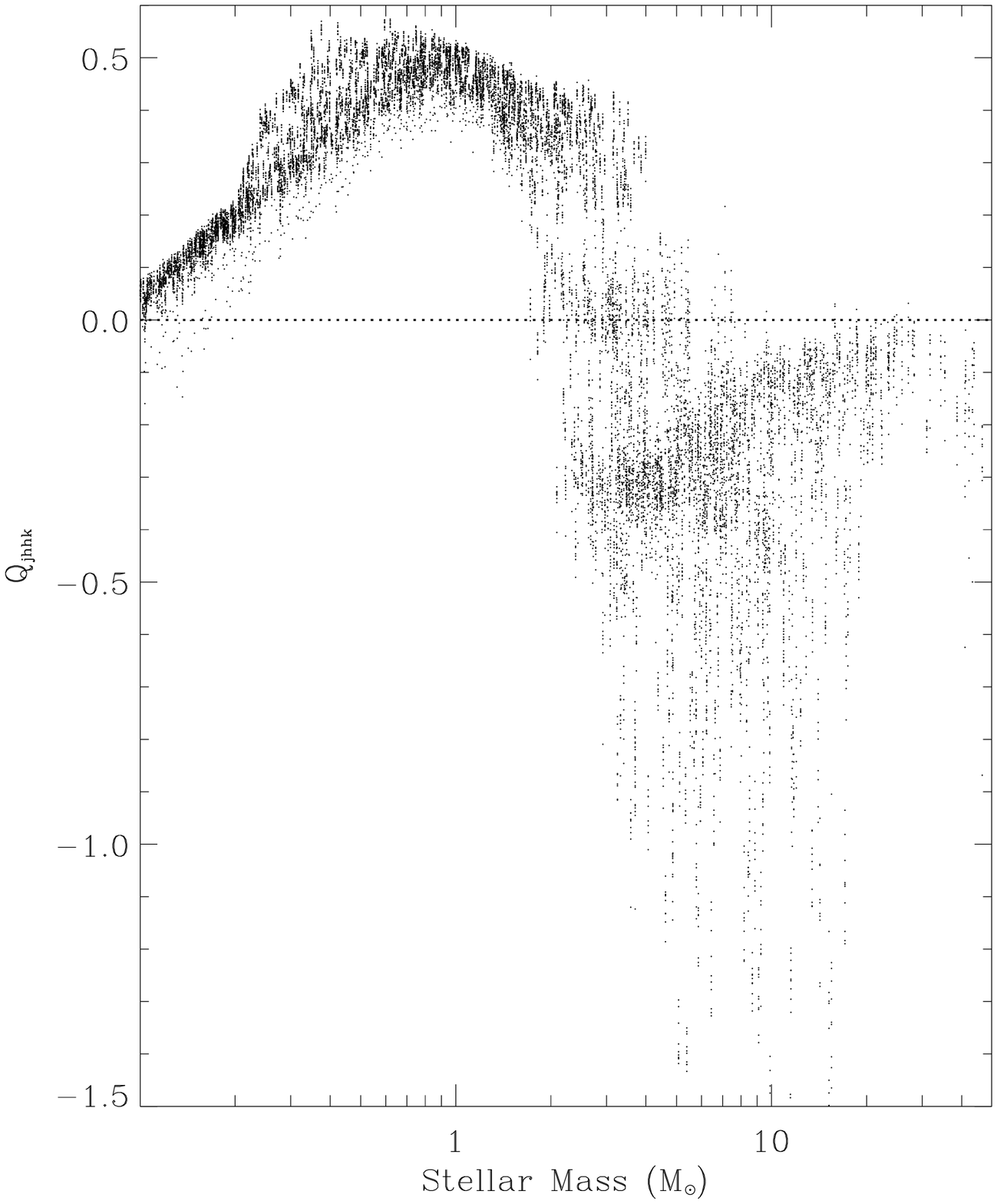}
	\includegraphics[width=5cm]{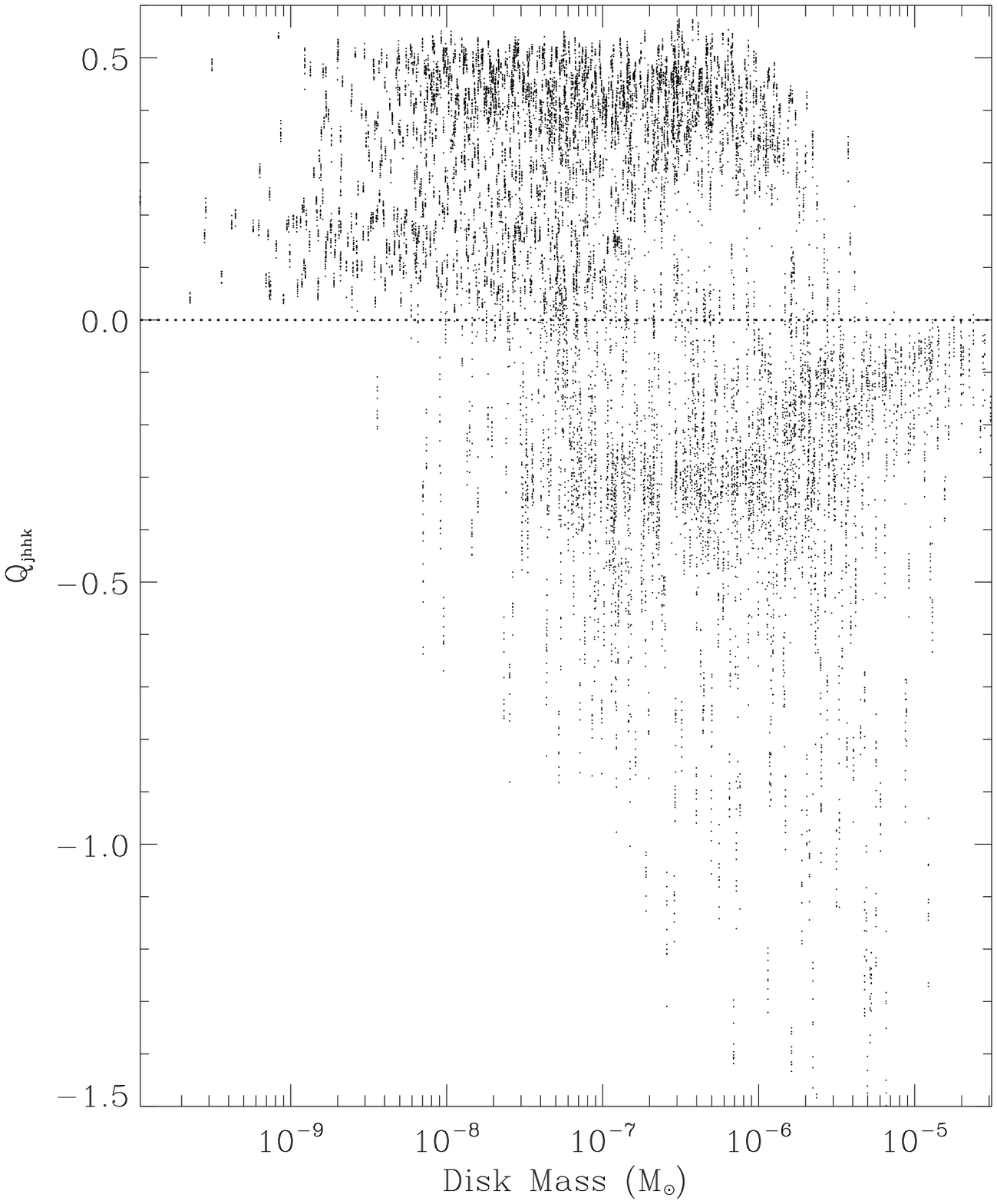}
	\caption{Diagrams of $Q_{JHHK}$ evaluated from the models of \citet{Robi06} of ClassII stars with thin disk, vs. the mass of the central star (upper panel) and the disk mass (lower panel). The dotted line is the limit of photospheric emission in the  $Q_{JHHK}$ index.}
	\label{qjhhkrobifig}
	\end{figure}

To verify the hypothesis that the adopted $Q$ indices can select objects with low-mass disks, we used the NIR photometry, in the 2MASS system, predicted by the models of \citet{Robi06}, to compute the expected $Q_{JHHK}$ index of disk-bearing objects. We evaluated this index for models of ClassII YSOs with thin and evolved disk, i.e. for which $M_{disk}/M_{star}\leq10^{-6}$. Fig. \ref{qjhhkrobifig} shows the diagrams of their $Q_{JHHK}$ vs. the mass of the central star (upper panel) and the mass of the disk (lower panel). In both panels several models have $Q_{JHHK}$ below the limit of photospheric emission, and then they would be selected with this index. The lower panel shows that most of these stars have disks with very low masses (from $10^{-5}$ to $10^{-8}$ $M_{\odot}$). For even lower masses the disk emission becomes too faint to give a negative $Q_{JHHK}$. In the upper panel, it is shown that negative indices are obtained for stars with masses over $\sim3 M_{\odot}$, likely because less massive stars cannot heat the material in the disk sufficiently, so its emission remains weak. These diagrams confirm the effectiveness of $Q$ indices in detecting thin disks around intermediate- and high-mass stars. \par

\begin{figure*}[]
	\centering	
	\includegraphics[width=7cm]{20515fit.eps}
	\includegraphics[width=7cm]{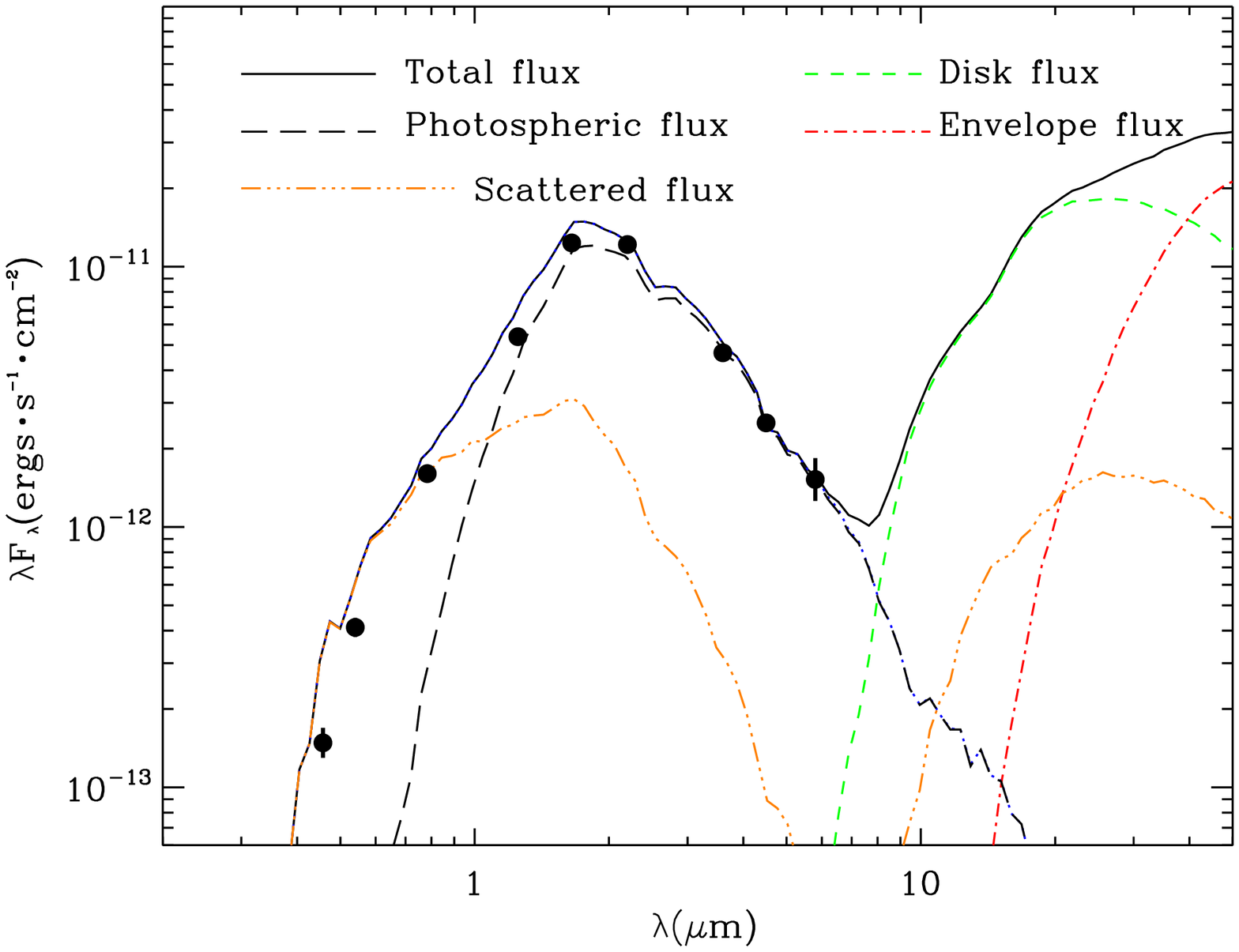}
	\caption{The observed SED (points) of the star 6611-20515. In the left panel the 5 compatible models (lines) and the predicted photospheric flux (dashed line) are also shown. The right panel shows the components of the best-fit model. As in Fig. \ref{sedEnoE}, all the SEDs are normalized at the distance of 1 Kpc, and the observed SED in the right panel has been dereddened using the visual extinction estimated by the SED model ($A_V=2.24^m$).}
	\label{20515bfm}
	\end{figure*}

	 The SEDs of 11 stars cannot be reconciled with any of the considered hypotheses. For these stars, the SED analysis suggests that the excesses detected with various $Q$ indices are due to an alteration of the optical SED by scattered light, as for the BWE stars. In the following, we use the example of the source 6611-20515 (see Fig. \ref{20515bfm}) to illustrate this effect. As in the case of 6611-27750 (Fig. \ref{sedEnoE}), the optical flux of this star is dominated by scattered light and not by the directly observed photospheric emission.

	\begin{table}[!h]
	\centering
	\caption {Stellar and disk parameters of the source 6611-20515 obtained with the SED analysis.}
	\vspace{0.3cm}
	\begin{tabular}{cc}
	\hline
	\hline
	Parameter & Value \\
	\hline
	\hline
	Age (Myrs)			&$\leq 0.2$\\
	$M_{\star}$ ($M_{\odot}$)	&$\leq 1 $ \\
	$M_{disk}$ ($M_{\odot}$)	&$1\times 10^{-5} \div 1\times 10^{-2}$\\
	$R_{in}$ (AU)			&$\geq 10$\\
	Inclination (degree)		&$30 \div 60$\\
	$\dot{M}$ ($M_{\odot}/yr$)	&$10^{-7} \div 10^{-9}$\\
	\hline
	\hline
	\end{tabular}
	\label{20855par}
	\end{table}

	The parameters of 6611-20515 obtained with SED analysis are summarized in Table \ref{20855par}. The excesses of 6611-20515 are detected using $Q_{2MASS}$, $Q_{UKIDSS}$ and, despite the normal $[3.6]-[4.5]$ color (equal to 0.08), $Q_{VIJ[3.6]}$ and $Q_{VIJ[4.5]}$ indices. The left panel of Fig. \ref{20515bfm} shows the observed SED of this star, with the 5 compatible model SEDs overplotted, while the right panel shows the components (i.e. the emission from star and disk) of the best compatible model. This model predicts a large inner disk radius equal to 26 AU, with the disk emission becoming significant only for $\lambda \geq 10 \mu m$, not accessible with our data, and that of the envelope at even longer wavelengths. The flux at shorter wavelengths (comprising all the bands in which the excesses have been detected) is emitted only by the photosphere, according to the observed value of $[3.6]-[4.5]$. However, in the right panel of Fig. \ref{20515bfm}, it is evident that the observed stellar flux is composed of both direct and scattered fluxes, as for the BWE stars. In particular, the optical/NIR emission is dominated by scattered light (more than 90\%) for $\lambda < 1 \mu m$ and by direct emission at longer wavelengths. The boundary between the regions of the spectrum dominated respectively by direct and scattered flux is at about $1 \mu m$, between WFI I band ($0.8 \mu m$) and J band ($1.2 \mu m$). As a consequence, the non photospheric $Q_{VIJ[3.6]}$ and $Q_{VIJ[4.5]}$ indices are produced by the different paths of the photospheric flux at these wavelengths, since $J-[3.6]$ is at a photospheric value, while $V-I$ is bluer due to the scattered light. The excesses detected in this star, then, are not related to the emission from the inner disk (that is evacuated) but they are due to the scatter of a significant amount of stellar light into the line of sight by disk and envelope. A further confirmation of the emission in JHK bands of 6611-20515 not coming from the disk is that its $Q_{JHHK}$ index, which is defined without using any optical band, is entirely compatible with a reddened photosphere. Finally, in the left panel of Fig. \ref{20515bfm} it is evident that the expected photospheric flux is larger than the observed one, since the disk is so inclined to partially occult the central star. We conclude that this star, and the other 11 sharing the same characteristics, does not show any NIR excesses in the IRAC and bands, but instead a $V$ excess induced by scattered light. \par

\section{Summary}
\label{thatsallfolks}

	In this paper we examine the SEDs of two samples of peculiar stars with excesses in 2MASS, UKIDSS, and IRAC bands, which were selected in our previous papers devoted to NGC~6611 and M16. The excesses were detected using suitable reddening-free color indices.   \par
The first sample is composed of 90 stars with excesses in infrared colors and with $V-I$ bluer than the cluster PMS locus in the $V$ vs. $V-I$ optical diagram. Thus, they are apparently older than the adopted upper limit of the cluster age. We discuss photometric and spectroscopic evidence supporting the membership of these stars to NGC~6611. SED analysis suggests that their optical colors are altered by a significant amount of light scattered by circumstellar material into the line of sight, which rises the observed optical flux in $V$ band. We present hints that this effect is more evident in young stars (younger than $10^6$ years) probably as a result of a higher density of small grains in the disk atmosphere than in disks around older stars. In several cases, a partial occultation of the stellar photosphere by an inclined disk, alone or together with scattering effects, can be invoked. The occultation, in fact, decreases the flux observed from the sources, but the optical colors remain unchanged. We conclude that the effects related to the presence of a disk (mostly if it is observed at high inclination) shift the position of disk-bearing stars in the $V$ vs. $V-I$ diagram from the PMS to the Main Sequence locus, altering the estimate of stellar mass, effective temperature, and age.\par
	The second sample of stars is composed of 66 stars with NIR excesses and photospheric IRAC colors. Our analysis suggests that these stars are disk-bearing members with
\begin{itemize}
\item a disk with a large inner hole and a broad silicate band in emission: in these cases the emission from the disk becomes significant only at 8.0$\mu m$ and at longer wavelengths;
\item or a disk with a low mass (i.e. with $M_{disk}/M_{star} \leq 10^{-6}$) producing small excesses in NIR bands. These excesses are not evident in color-magnitude diagrams, in which these stars can be easily confused with normal colors stars, but they are selected by our peculiar disk diagnostic, which compares infrared and optical colors;
\item or a disk that significantly scatters the stellar optical light into the line of sight. These stars are not selected thanks to excesses in infrared bands induced by the disk, but thanks to the light scattered by the disk, which affects the optical colors.
\end{itemize}


\begin{acknowledgements}
Support for this work has been provided by the CONSTELLATION grant YA 2007 and the contract PRIN-INAF (P.I.: Lanza). This work is based on data obtained as part of the UKIRT Infrared Deep Sky Survey, public data obtained with WFI@ESO, 2MASS Point Source Catalog and GLIMPSE survey with Spitzer/IRAC. The authors acknowledge M. Caramazza for her help in Spitzer data analysis.
\end{acknowledgements}

\newpage
\addcontentsline{toc}{section}{\bf Bibliografia}
\bibliographystyle{apj}
\bibliography{biblio}

\begin{thebibliography}{34}
\expandafter\ifx\csname natexlab\endcsname\relax\def\natexlab#1{#1}\fi

\bibitem[{{Argiroffi} {et~al.}(2007){Argiroffi}, {Maggio}, \& {Peres}}]{Argi07}
{Argiroffi}, C., {Maggio}, A., \& {Peres}, G. 2007, \aap, 465, L5

\bibitem[{{Benjamin} {et~al.}(2003){Benjamin}, {Churchwell}, {Babler}, {Bania},
  {Clemens}, {Cohen}, {Dickey}, {Indebetouw}, {Jackson}, {Kobulnicky},
  {Lazarian}, {Marston}, {Mathis}, {Meade}, {Seager}, {Stolovy}, {Watson},
  {Whitney}, {Wolff}, \& {Wolfire}}]{Ben03}
{Benjamin}, R.~A., {Churchwell}, E., {Babler}, B.~L., {Bania}, T.~M.,
  {Clemens}, D.~P., {Cohen}, M., {Dickey}, J.~M., {Indebetouw}, R., {Jackson},
  J.~M., {Kobulnicky}, H.~A., {Lazarian}, A., {Marston}, A.~P., {Mathis},
  J.~S., {Meade}, M.~R., {Seager}, S., {Stolovy}, S.~R., {Watson}, C.,
  {Whitney}, B.~A., {Wolff}, M.~J., \& {Wolfire}, M.~G. 2003, \pasp, 115, 953

\bibitem[{{Carey} {et~al.}(2009){Carey}, {Noriega-Crespo}, {Mizuno}, {Shenoy},
  {Paladini}, {Kraemer}, {Price}, {Flagey}, {Ryan}, {Ingalls}, {Kuchar},
  {Pinheiro Gon{\c c}alves}, {Indebetouw}, {Billot}, {Marleau}, {Padgett},
  {Rebull}, {Bressert}, {Ali}, {Molinari}, {Martin}, {Berriman}, {Boulanger},
  {Latter}, {Miville-Deschenes}, {Shipman}, \& {Testi}}]{Car09}
{Carey}, S.~J., {Noriega-Crespo}, A., {Mizuno}, D.~R., {Shenoy}, S.,
  {Paladini}, R., {Kraemer}, K.~E., {Price}, S.~D., {Flagey}, N., {Ryan}, E.,
  {Ingalls}, J.~G., {Kuchar}, T.~A., {Pinheiro Gon{\c c}alves}, D.,
  {Indebetouw}, R., {Billot}, N., {Marleau}, F.~R., {Padgett}, D.~L., {Rebull},
  L.~M., {Bressert}, E., {Ali}, B., {Molinari}, S., {Martin}, P.~G.,
  {Berriman}, G.~B., {Boulanger}, F., {Latter}, W.~B., {Miville-Deschenes},
  M.~A., {Shipman}, R., \& {Testi}, L. 2009, \pasp, 121, 76

\bibitem[{{Cutri} {et~al.}(2003){Cutri}, {Skrutskie}, {van Dyk}, {Beichman},
  {Carpenter}, {Chester}, {Cambresy}, {Evans}, {Fowler}, {Gizis}, {Howard},
  {Huchra}, {Jarrett}, {Kopan}, {Kirkpatrick}, {Light}, {Marsh}, {McCallon},
  {Schneider}, {Stiening}, {Sykes}, {Weinberg}, {Wheaton}, {Wheelock}, \&
  {Zacarias}}]{Cutri03}
{Cutri}, R.~M., {Skrutskie}, M.~F., {van Dyk}, S., {Beichman}, C.~A.,
  {Carpenter}, J.~M., {Chester}, T., {Cambresy}, L., {Evans}, T., {Fowler}, J.,
  {Gizis}, J., {Howard}, E., {Huchra}, J., {Jarrett}, T., {Kopan}, E.~L.,
  {Kirkpatrick}, J.~D., {Light}, R.~M., {Marsh}, K.~A., {McCallon}, H.,
  {Schneider}, S., {Stiening}, R., {Sykes}, M., {Weinberg}, M., {Wheaton},
  W.~A., {Wheelock}, S., \& {Zacarias}, N. 2003, VizieR Online Data Catalog,
  2246, 0

\bibitem[{{D'Alessio} {et~al.}(2001){D'Alessio}, {Calvet}, \&
  {Hartmann}}]{dale01}
{D'Alessio}, P., {Calvet}, N., \& {Hartmann}, L. 2001, \apj, 553, 321

\bibitem[{{D'Alessio} {et~al.}(1999){D'Alessio}, {Calvet}, {Hartmann},
  {Lizano}, \& {Cant{\'o}}}]{dale99}
{D'Alessio}, P., {Calvet}, N., {Hartmann}, L., {Lizano}, S., \& {Cant{\'o}}, J.
  1999, \apj, 527, 893

\bibitem[{{D'Alessio} {et~al.}(1998){D'Alessio}, {Canto}, {Calvet}, \&
  {Lizano}}]{dale98}
{D'Alessio}, P., {Canto}, J., {Calvet}, N., \& {Lizano}, S. 1998, \apj, 500,
  411

\bibitem[{{D'Alessio} {et~al.}(2005){D'Alessio}, {Mer{\'{\i}}n}, {Calvet},
  {Hartmann}, \& {Montesinos}}]{dale05}
{D'Alessio}, P., {Mer{\'{\i}}n}, B., {Calvet}, N., {Hartmann}, L., \&
  {Montesinos}, B. 2005, Revista Mexicana de Astronomia y Astrofisica, 41, 61

\bibitem[{{Damiani} {et~al.}(2006){Damiani}, {Prisinzano}, {Micela}, \&
  {Sciortino}}]{Dami06}
{Damiani}, F., {Prisinzano}, L., {Micela}, G., \& {Sciortino}, S. 2006, \aap,
  459, 477

\bibitem[{{Dullemond} \& {Dominik}(2004)}]{DD04B}
{Dullemond}, C.~P. \& {Dominik}, C. 2004, \aap, 421, 1075

\bibitem[{{Flagey} {et~al.}(2009){Flagey}, {Boulanger}, {Noriega-Crespo},
  {Carey}, \& {Mizuno}}]{Fla09}
{Flagey}, N., {Boulanger}, F., {Noriega-Crespo}, A., {Carey}, S., \& {Mizuno},
  D. 2009, in The Evolving ISM in the Milky Way and Nearby Galaxies

\bibitem[{{Flaherty} {et~al.}(2007){Flaherty}, {Pipher}, {Megeath}, {Winston},
  {Gutermuth}, {Muzerolle}, \& {Fazio}}]{Fla07}
{Flaherty}, K.~M., {Pipher}, J.~L., {Megeath}, S.~T., {Winston}, E.~M.,
  {Gutermuth}, R.~A., {Muzerolle}, J., \& {Fazio}, G.~G. 2007, ArXiv
  Astrophysics e-prints

\bibitem[{{Guarcello} {et~al.}(2009){Guarcello}, {Micela}, {Damiani}, {Peres},
  {Prisinzano}, \& {Sciortino}}]{Io09}
{Guarcello}, M.~G., {Micela}, G., {Damiani}, F., {Peres}, G., {Prisinzano}, L.,
  \& {Sciortino}, S. 2009, \aap, 496, 453

\bibitem[{{Guarcello} {et~al.}(2010){Guarcello}, {Micela}, {Peres},
  {Prisinzano}, \& {Sciortino}}]{Io10}
{Guarcello}, M.~G., {Micela}, G., {Peres}, G., {Prisinzano}, L., \&
  {Sciortino}, S. 2010, in preparation

\bibitem[{{Guarcello} {et~al.}(2007){Guarcello}, {Prisinzano}, {Micela},
  {Damiani}, {Peres}, \& {Sciortino}}]{io07}
{Guarcello}, M.~G., {Prisinzano}, L., {Micela}, G., {Damiani}, F., {Peres}, G.,
  \& {Sciortino}, S. 2007, \aap, 462, 245

\bibitem[{{Hartmann} \& {Kenyon}(1990)}]{Har90}
{Hartmann}, L.~W. \& {Kenyon}, S.~J. 1990, \apj, 349, 190

\bibitem[{{Hillenbrand}(1997)}]{Hille97}
{Hillenbrand}, L.~A. 1997, \aj, 113, 1733

\bibitem[{{Kurosawa} {et~al.}(2006){Kurosawa}, {Harries}, \&
  {Symington}}]{Kuro06}
{Kurosawa}, R., {Harries}, T.~J., \& {Symington}, N.~H. 2006, \mnras, 370, 580

\bibitem[{{Lucas} {et~al.}(2008){Lucas}, {Hoare}, {Longmore}, {Schr{\"o}der},
  {Davis}, {Adamson}, {Bandyopadhyay}, {de Grijs}, {Smith}, {Gosling},
  {Mitchison}, {G{\'a}sp{\'a}r}, {Coe}, {Tamura}, {Parker}, {Irwin}, {Hambly},
  {Bryant}, {Collins}, {Cross}, {Evans}, {Gonzalez-Solares}, {Hodgkin},
  {Lewis}, {Read}, {Riello}, {Sutorius}, {Lawrence}, {Drew}, {Dye}, \&
  {Thompson}}]{Luca08}
{Lucas}, P.~W., {Hoare}, M.~G., {Longmore}, A., {Schr{\"o}der}, A.~C., {Davis},
  C.~J., {Adamson}, A., {Bandyopadhyay}, R.~M., {de Grijs}, R., {Smith}, M.,
  {Gosling}, A., {Mitchison}, S., {G{\'a}sp{\'a}r}, A., {Coe}, M., {Tamura},
  M., {Parker}, Q., {Irwin}, M., {Hambly}, N., {Bryant}, J., {Collins}, R.~S.,
  {Cross}, N., {Evans}, D.~W., {Gonzalez-Solares}, E., {Hodgkin}, S., {Lewis},
  J., {Read}, M., {Riello}, M., {Sutorius}, E.~T.~W., {Lawrence}, A., {Drew},
  J.~E., {Dye}, S., \& {Thompson}, M.~A. 2008, \mnras, 391, 136

\bibitem[{{Mathis}(1990)}]{Mat90}
{Mathis}, J.~S. 1990, \araa, 28, 37

\bibitem[{{Megeath} {et~al.}(2004){Megeath}, {Allen}, {Gutermuth}, {Pipher},
  {Myers}, {Calvet}, {Hartmann}, {Muzerolle}, \& {Fazio}}]{Mege04}
{Megeath}, S.~T., {Allen}, L.~E., {Gutermuth}, R.~A., {Pipher}, J.~L., {Myers},
  P.~C., {Calvet}, N., {Hartmann}, L., {Muzerolle}, J., \& {Fazio}, G.~G. 2004,
  \apjs, 154, 367

\bibitem[{{Momany} {et~al.}(2001){Momany}, {Vandame}, {Zaggia}, {Mignani}, {da
  Costa}, {Arnouts}, {Groenewegen}, {Hatziminaoglou}, {Madejsky}, {Rit{\'e}},
  {Schirmer}, \& {Slijkhuis}}]{Moma01}
{Momany}, Y., {Vandame}, B., {Zaggia}, S., {Mignani}, R.~P., {da Costa}, L.,
  {Arnouts}, S., {Groenewegen}, M.~A.~T., {Hatziminaoglou}, E., {Madejsky}, R.,
  {Rit{\'e}}, C., {Schirmer}, M., \& {Slijkhuis}, R. 2001, \aap, 379, 436

\bibitem[{{Munari} \& {Carraro}(1996)}]{Muna96}
{Munari}, U. \& {Carraro}, G. 1996, \aap, 314, 108

\bibitem[{{Oliveira}(2008)}]{Oli08}
{Oliveira}, J.~M. 2008, {Star Formation in the Eagle Nebula}, ed.
  B.~{Reipurth}, 599--+

\bibitem[{{Padgett} {et~al.}(1999){Padgett}, {Brandner}, {Stapelfeldt},
  {Strom}, {Terebey}, \& {Koerner}}]{Pad99}
{Padgett}, D.~L., {Brandner}, W., {Stapelfeldt}, K.~R., {Strom}, S.~E.,
  {Terebey}, S., \& {Koerner}, D. 1999, \aj, 117, 1490

\bibitem[{{Palla} {et~al.}(2005){Palla}, {Randich}, {Flaccomio}, \&
  {Pallavicini}}]{Pal05}
{Palla}, F., {Randich}, S., {Flaccomio}, E., \& {Pallavicini}, R. 2005, \apjl,
  626, L49

\bibitem[{{Palla} {et~al.}(2007){Palla}, {Randich}, {Pavlenko}, {Flaccomio}, \&
  {Pallavicini}}]{Pa07}
{Palla}, F., {Randich}, S., {Pavlenko}, Y.~V., {Flaccomio}, E., \&
  {Pallavicini}, R. 2007, \apjl, 659, L41

\bibitem[{{Pringle}(1981)}]{Prin81}
{Pringle}, J.~E. 1981, \araa, 19, 137

\bibitem[{{Rieke} \& {Lebofsky}(1985)}]{RL85}
{Rieke}, G.~H. \& {Lebofsky}, M.~J. 1985, \apj, 288, 618

\bibitem[{{Robitaille} {et~al.}(2007){Robitaille}, {Whitney}, {Indebetouw}, \&
  {Wood}}]{Robi07}
{Robitaille}, T.~P., {Whitney}, B.~A., {Indebetouw}, R., \& {Wood}, K. 2007,
  \apjs, 169, 328

\bibitem[{{Robitaille} {et~al.}(2006){Robitaille}, {Whitney}, {Indebetouw},
  {Wood}, \& {Denzmore}}]{Robi06}
{Robitaille}, T.~P., {Whitney}, B.~A., {Indebetouw}, R., {Wood}, K., \&
  {Denzmore}, P. 2006, \apjs, 167, 256

\bibitem[{{Shakura} {et~al.}(1978){Shakura}, {Sunyaev}, \&
  {Zilitinkevich}}]{SS78}
{Shakura}, N.~I., {Sunyaev}, R.~A., \& {Zilitinkevich}, S.~S. 1978, \aap, 62,
  179

\bibitem[{{Siess} {et~al.}(2000){Siess}, {Dufour}, \& {Forestini}}]{Sie00}
{Siess}, L., {Dufour}, E., \& {Forestini}, M. 2000, \aap, 358, 593

\bibitem[{{Throop} {et~al.}(2001){Throop}, {Bally}, {Esposito}, \&
  {McCaughrean}}]{Thro01}
{Throop}, H.~B., {Bally}, J., {Esposito}, L.~W., \& {McCaughrean}, M.~J. 2001,
  Science, 292, 1686

\end{thebibliography}
\end{document}